# Developing an improved Crystal Graph Convolutional Neural Network framework for accelerated materials discovery

Cheol Woo Park, Chris Wolverton*

*Department of Materials Science and Engineering, Northwestern University, Evanston, Illinois 60208, USA*

E-mail: c-wolverton@northwestern.edu

## Abstract

The recently proposed crystal graph convolutional neural network (CGCNN) offers a highly versatile and accurate machine learning (ML) framework by learning material properties directly from graph-like representations of crystal structures ("crystal graphs"). Here, we develop an improved variant of the CGCNN model (iCGCNN) that outperforms the original by incorporating information of the Voronoi tessellated crystal structure, explicit 3-body correlations of neighboring constituent atoms, and an optimized chemical representation of interatomic bonds in the crystal graphs. We demonstrate the accuracy of the improved framework in two distinct illustrations: First, when trained/validated on 180,000/20,000 density functional theory (DFT) calculated thermodynamic stability entries taken from the Open Quantum Materials Database (OQMD) and evaluated on a separate test set of 230,000 entries, iCGCNN achieves a predictive accuracy that is significantly improved, i.e., 20% higher than that of the original CGCNN. Second, when used to assist high-throughput search for materials in the $ThCr_2Si_2$ structure-type, iCGCNN exhibited a success rate of 31% which is 310 times higher than an undirected high-throughput search and 2.4 times higher than that of the original CGCNN. Using both CGCNN and iCGCNN, we screened 132,600 compounds with elemental decorations of the $ThCr_2Si_2$ prototype crystal structure and identified a total of 97 new unique stable compounds by performing 757 DFT calculations, accelerating the computational time of the high-throughput search by a factor of 130. Our results suggest that the iCGCNN can be used to accelerate high-throughput discoveries of new materials by quickly and accurately identifying crystalline compounds with properties of interest.

**Introduction**

Density functional theory (DFT) calculations have proven to be a valuable tool in characterizing materials properties and discovering new materials [1]. However, prediction of novel materials through DFT calculations remains a computationally challenging process due to the sheer size of the materials search space. Recently, with the availability of large material databases [2–7], data-driven materials design and discovery using machine learning (ML) has gained much attention for its potential to predict new materials with favorable properties much faster than DFT calculations with substantially less computational cost. ML models have been developed for various materials applications such as predicting formation energies [14–18, 31], band gap energies [19–23], melting temperatures [24-26], thermal conductivity [25, 27], and mechanical properties of materials [28-30].

A working ML model requires three components:1) training and testing data, 2) a ML algorithm, and 3) materials representation. Much of the creative efforts in materials informatics have been focused on developing representations that can uniquely define each material and best capture the chemistry that influences the property of interest. Recently, inspired by the breakthroughs made in other fields such as computer vision, there has been a rising effort to take advantage of neural networks to extract useful descriptors from inorganic compounds without having to construct them manually [32-38, 45]. In particular, graph neural networks (GNNs), first used by the quantum chemistry community to extract descriptors from molecular graphs [32-36], have started being used on graph representations of crystal compounds to reach unprecedented accuracy in predicting materials properties and to gain chemical insight [36-38, 48].

In this work, we show that frameworks utilizing GNNs can be further improved in predicting material properties. We build upon the recently proposed crystal graph convolution neural network (CGCNN) framework [37] which utilizes graph representations of crystals, referred to as crystal graphs, and present an improved framework (iCGCNN) that better represents the chemical nature of an inorganic compound. In this new framework, descriptors extracted from the crystal graphs include the information of the Voronoi tessellated crystal structure, explicit 3-body correlations of neighboring constituent atoms, and an optimized chemical representation of interatomic bonds, all of which are absent in the crystal graphs utilized by the original framework. The improvement of the new model is illustrated through two distinct tests.

First, we compare the accuracy of CGCNN and iCGCNN in predicting the thermodynamic stability of inorganic materials, using a training/testing dataset of DFT-calculated stabilities from the Open Quantum Materials Database (OQMD) [2, 3]. Thermodynamic stability in this work refers to the difference between the formation energy of a compound and the lowest-energy linear combination of phases corresponding to that composition, typically calculated from the so-called convex hull constructions (henceforth, "convex hull energy"). We use two different approaches to predict stability. In the first approach, we train ML models to predict the formation energy of phases, which is subsequently used to calculate stability relative to the convex hull energy derived from DFT calculated formation energies. In the second approach, we train and test the ML models directly on the DFT-calculated thermodynamic stability data, bypassing calculations of formation energy.

In the second illustration, we conduct separate ML-assisted high-throughput searches using both CGCNN and iCGCNN to discover new stable compounds in the $ThCr_2Si_2$ structure-type, one of the most commonly occurring ternary prototype structures. We compare the performances of CGCNN and iCGCNN based on the number compounds that were confirmed to be stable through DFT and the success rate, which we define as the ratio of number of stable compounds identified to the number of DFT calculations that were performed to identify those stable compounds, in their respective high-throughput search.

In both studies, we find that iCGCNN significantly outperforms the original CGCNN model. In predicting thermodynamic stability, iCGCNN achieves an accuracy that is 20% higher than that of CGCNN. In predicting new stable $ThCr_2Si_2$-type materials, iCGCNN identifies nearly twice as many compounds with a success rate that is greater by a factor of 2.4 than the original CGCNN. Using both CGCNN and iCGCNN, we screened 132,600 $ThCr_2Si_2$-type compounds that were generated by substituting elements into the original $ThCr_2Si_2$ structure and identified 97 of them to be stable by only performing 757 DFT-calculations, a success rate that is higher than that of an undirected high-throughput search by a factor of 130. Our findings show iCGCNN to be a highly efficient screening tool to predict potentially stable materials to accelerate the challenging task of materials design and discovery.

## Results
**Description of improvements in the iCGCNN**

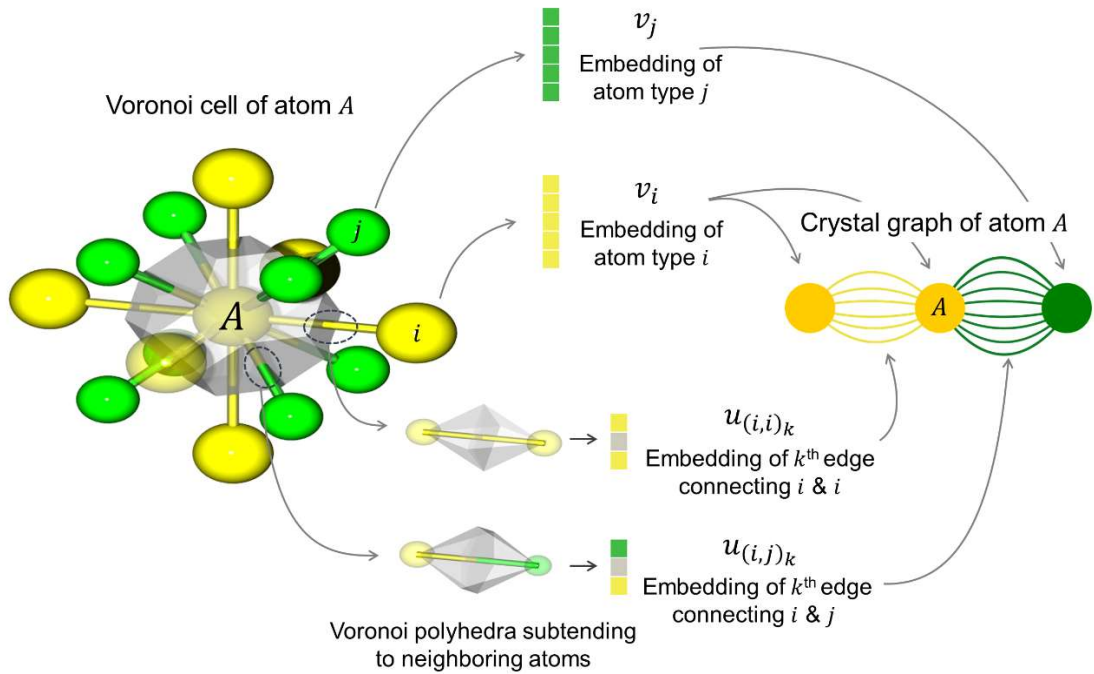

**Figure 1** Illustration of iCGCNN crystal graph. The crystal graph shown on the far right represents the local environment of atom $A$. Multiple edges connect $A$ to neighboring nodes to show the number of Voronoi neighbors. The nodes and edges are embedded with vectors that characterize the constituent atoms ($v_i$, $v_j$) and their correlations with neighboring atoms ($u_{(i,i)_k}$, $u_{(i,j)_k}$) respectively. Edge vectors include information about the Voronoi polyhedra such as solid angle, area, and volume.

The original CGCNN framework [37] utilizes a graph representation of the crystals that is composed of two parts: 1) nodes that represent constituent atoms of the crystal, and 2) edges that represent the bonds between neighboring atoms. A node is embedded with a vector $\boldsymbol{v}_i$ to represent the properties of atom $i$, where we define embedding as the process of mapping a discrete object to a vector of real numbers. Each edge is also embedded with a vector $\boldsymbol{u}_{(i,j)_k}$ that contains the distance information between neighboring atoms $i$ and $j$ of the crystal unit cell. In order to account for the periodicity of the crystal, multiple edges between atoms $i$ and $j$, as indexed by $k$, can exist. Each node in the crystal graph is connected to its 12 nearest neighbors. During the training phase, the node vectors are updated iteratively according to a convolution function defined by Equation 1: $v_i^{(t+1)} = v_i^{(t)} + \sum_{j,k} \sigma\left(z_{(i,j)_k}^{(t)} W_1^{(t)} + b_1^{(t)}\right) \odot g\left(z_{(i,j)_k}^{(t)} W_2^{(t)} + b_2^{(t)}\right)$

The terms in the sum operator represent the 2-body correlation of an atom with its neighboring atoms. $z_{(i,j)_k}^{(t)} = v_i^{(t)} \oplus v_j^{(t)} \oplus u_{(i,j)_k}^{(t)}$ is the concatenation of the node and edge vectors. $\odot$ represents a matrix element-wise multiplication while $\sigma$ and $g$ respectively represent a sigmoid function and a nonlinear activation function. $W^{(t)}$ and $b^{(t)}$ represent the weight and bias matrices respectively for the $t$-th convolution step. A more detailed explanation of the CGCNN framework can be found in Ref [37].

CGCNN offers a highly flexible framework to represent different crystal structures and exhibits excellent performances in predicting a variety of material properties. However, the design of the crystal graphs utilized by CGCNN may not be optimal in representing the chemical environment of constituent atoms. We identify 3 possible drawbacks of the original CGCNN model that we attempt to address in the new iCGCNN model.

The first drawback of CGCNN is that regardless of the crystal structure being represented, each node in every crystal graph is connected to 12 of its nearest neighbors. While the local chemical environment of an atom is determined by all of its neighboring atoms, atoms in the first and second nearest neighbor shell often have the largest impact on the local environment. Depending on the crystal structure, it is possible that the 12 nearest neighbors of an atom may include neighbors beyond the first and second nearest neighbor shell. These neighbors may introduce information that could overshadow the more important information relayed from the nearest neighbors and deter the ML model from learning the most optimal local environment representation of an atom during training. In our improved model, we attempt to better represent the local environment of crystals by connecting each node to its Voronoi neighbors, as illustrated in Figure 1. Furthermore, such a construction enables us to use information from the Voronoi tessellation of the crystal [31] in the edge vector embeddings of the crystal graph, in addition to interatomic distance information. Voronoi polyhedral information embedded in the edge vectors in the iCGCNN model include attributes such as the solid angle, area, and volume of the Voronoi cell subtended for the facet as calculated by the open source library, pymatgen [43].

The second drawback of the CGCNN is that only pair-wise correlations are explicitly encoded into the convolution function. Higher-order, many-body correlations (e.g., 3-body) are not explicitly encoded. We note that the original CGCNN implicitly encodes some information about the many-body correlations into the node vectors through multiple iterations of the convolution step. However, much of the information regarding the many-body correlations is inevitably lost as it is not explicitly encoded into convolutional function. To minimize the loss of information, we *explicitly* integrate information of 3-body interactions between atoms $i, j$, and $l$ into the convolution function in the iCGCNN model by adding the following term to equation (1):

Equation 2: $\sum_{j,l,k,k'} \sigma\left(z'^{(t)}_{(i,j,l)_{k,k'}} W'^{(t)}_1 + b'^{(t)}_1\right) \odot g\left(z'^{(t)}_{(i,j,l)_{k,k'}} W'^{(t)}_2 + b'^{(t)}_2\right)$

The node vectors and edge vectors that connect atoms $i, j$, and $l$ are concatenated to form
$z'^{(t)}_{(i,j,l)_{k,k'}} = v^{(t)}_i \oplus v^{(t)}_j \oplus v^{(t)}_l \oplus u^{(t)}_{(i,j)_k} \oplus u^{(t)}_{(i,l)_{k'}}$.

The third drawback of the CGCNN is that the chemical representations of interatomic bonds, as defined by the edge vectors, are not optimized. In the original CGCNN, node vectors are iteratively optimized to better represent the local chemical environment of an atom, but the edge vectors remain unchanged during training. Thus, in CGCNN, interatomic bonds are not as well represented by the edge vectors as the local chemical environments are represented by the node vectors. To address this issue, we implemented the following convolutional function to update the edge vectors in iCGCNN,

Equation 3:

$$\boldsymbol{u}_{(i,j)_k}^{(t+1)} = \boldsymbol{u}_{(i,j)_k}^{(t)} + \sigma\left(\boldsymbol{z}_{(i,j)_k}^{(t)} \boldsymbol{W}_1^{(t)} + \boldsymbol{b}_1^{(t)}\right) \odot g\left(\boldsymbol{z}_{(i,j)_k}^{(t)} \boldsymbol{W}_2^{(t)} + \boldsymbol{b}_2^{(t)}\right)$$
$$+ \sum_{l,k\prime} \sigma\left(\boldsymbol{z'}_{(i,j,l)_{k,k\prime}}^{(t)} \boldsymbol{W'}_1^{(t)} + \boldsymbol{b'}_1^{(t)}\right) \odot g\left(\boldsymbol{z'}_{(i,j,l)_{k,k\prime}}^{(t)} \boldsymbol{W'}_2^{(t)} + \boldsymbol{b'}_2^{(t)}\right)$$

The terms in this convolutional function closely resemble those in equations (1) and (2). The terms in $\sigma\left(\boldsymbol{z}_{(i,j)_k}^{(t)} \boldsymbol{W}_1^{(t)} + \boldsymbol{b}_1^{(t)}\right) \odot g\left(\boldsymbol{z}_{(i,j)_k}^{(t)} \boldsymbol{W}_2^{(t)} + \boldsymbol{b}_2^{(t)}\right)$ represent how the chemical properties of the atoms $i$ and $j$ affect their chemical bond. The sum operator term represents how the interatomic bonds are affected by the presence of other nearby atoms.

**Predicting thermodynamic stability using CGCNN and iCGCNN**

In this section, we compare the predictive accuracies of the original and improved CGCNN model in predicting the thermodynamic stability of crystal compounds. The thermodynamic stability of a compound is determined by what is often referred to as "distance to convex hull" or simply "hull distance." The *convex hull* is defined as the envelope connecting the lowest energy compounds in the chemical space (e.g. binary Li-O space, ternary Cu-Mn-Al space). For example, in the binary Li-O chemical space, the convex hull is simply the envelope that connects the stable Li, $Li_2O$, $Li_2O_2$, $LiO_2$, $LiO_3$, and $O_2$ phases. The hull distance of a compound $i$, $\Delta H_{stab}^i$ is given by $\Delta H_{stab}^i = \Delta H_f^i - E_{hull}^i$, where $\Delta H_f^i$ is the formation energy of compound $i$ and $E_{hull}^i$ is the convex hull (*constructed without including i*) energy at the composition of $i$. Using the previous example, $\Delta H_{stab}^{Li_2O}$ is calculated by constructing a convex hull in the Li-O space after explicitly excluding $Li_2O$. It thus follows, as defined here, all stable compounds have a hull distance $\Delta H_{stab}^i \leq 0$ meV/atom. Note that our definition of hull distance is different from one of the common ways of defining it where $E_{hull}^i$ is the energy of the convex hull constructed including $i$. Under such a definition, stable compounds have a hull distance of zero.

In this study, we considered two strategies to determine the hull distances, using a combination of ML and DFT energies. In Strategy 1, we train the ML models on DFT-calculated formation energies of the compounds in the training dataset. Hull distances of the compounds in the test dataset are evaluated by taking the differences between the ML-predicted formation energies and DFT-calculated convex hull energies as in the OQMD. This approach requires us to calculate the convex hull energy every time we are predicting the hull distance of a new compound. In Strategy 2, we train the ML models directly on the DFT-calculated hull distances. This approach requires us to construct the convex hulls for compounds in the training data, but it allows us to bypass the convex hull construction altogether when predicting the stability of a new compound.

CGCNN and iCGCNN models using Strategy 1 were trained and validated on the formation energies of the ~200,000 compounds in the training and validation data taken from the OQMD. The models were then used to evaluate both the formation energies and hull distances of the ~230,000 compounds in the test data, where the hull distances are computed by taking the differences between the ML-predicted formation energies and the convex hull energies as calculated in the OQMD at that composition. Model performances utilizing Strategy 1 are summarized in Figure 2. Mean absolute errors (MAE) for formation energy predictions are 41.3 and 30.5 meV/atom for the CGCNN and iCGCNN respectively (Figure 2 (a) (b)). In comparison, Ward *et al*. report a MAE of 80 meV/atom in cross validation for the Voronoi tessellation model [31] when trained and tested on 435,000 formation energies taken from the OQMD. This shows that both CGCNN and iCGCNN offer highly accurate estimations of DFT-calculated formation energies compared to previously developed ML models, consistent to the results reported in Ref [37]. In another comparison, DFT is widely considered to be a reliable method in estimating various material properties where, for many cases, the differences between DFT and experimental results are trivial. For measuring formation energies, the difference between the DFT and experiments is around 100 meV/atom [3]. This implies that both CGCNN and iCGCNN can be used as a reliable method to estimate DFT-calculated material properties. MAE for hull distances are 41 and 30 meV/atom for CGCNN and iCGCNN respectively, identical to the formation energy prediction errors as expected since the hull distances are calculated directly from the formation energies (Figure 2 (c) (d)). Overall, iCGCNN hull distance prediction errors are lower than CGCNN by 25% and 15% in terms of MAE and root mean squared error (RMSE) respectively.

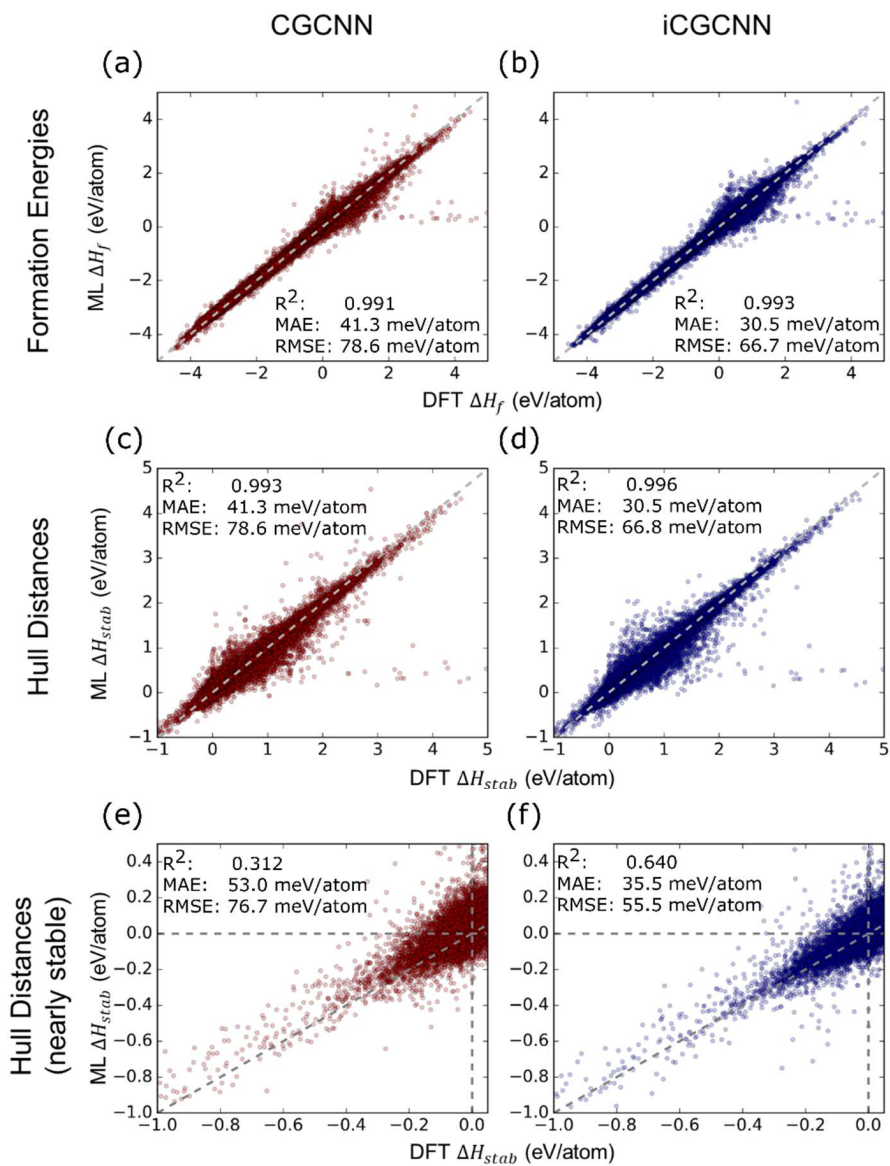

**Figure 2** Predictive accuracies of CGCNN and iCGCNN using Strategy 1. Hull distances are evaluated by taking the differences between ML-predicted formation energies and DFT-calculated convex hull energies. (a) (b) DFT vs ML formation energies for (a) CGCNN and (b) iCGCNN. (c) (d) DFT vs ML predicted hull distances for (c) CGCNN and (d) iCGCNN. Close up on compounds with hull distances smaller than 50 meV/atom for (c) and (d) are shown in (e) and (f) respectively.

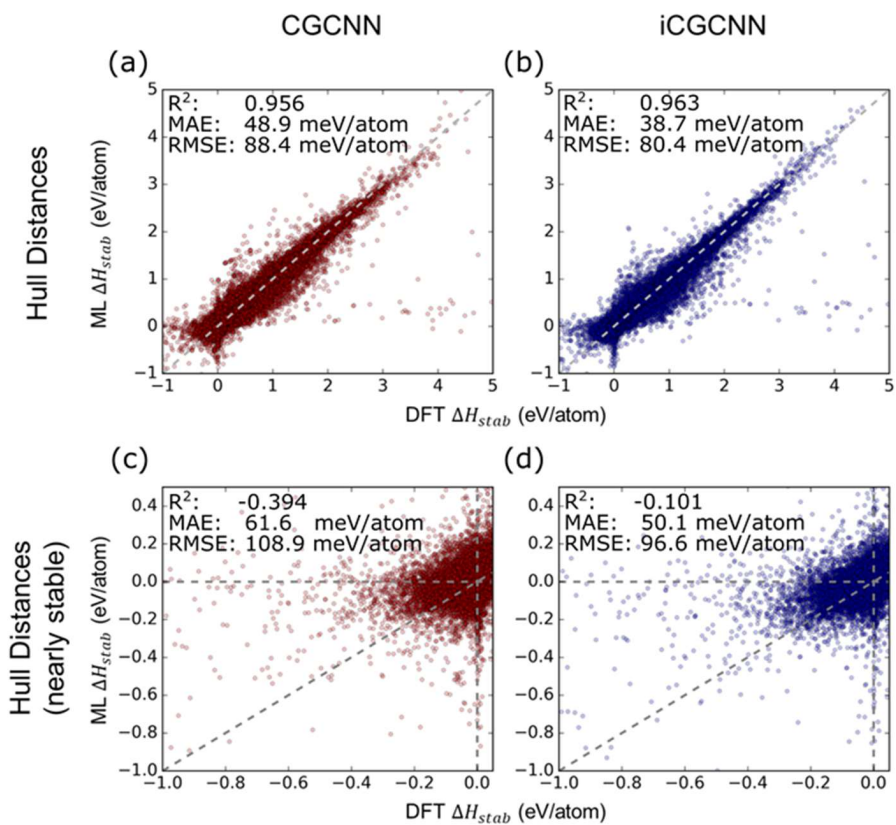

**Figure 3** Predictive accuracies of CGCNN and iCGCNN using Strategy 2. ML models are trained to directly predict the hull distances. (a) (b) DFT vs ML predicted hull distances for (a) CGCNN and (b) iCGCNN. Close-up on compounds with hull distances smaller than 50 meV/atom for (a) and (b) are shown in (c) and (d) respectively.

While it is important that a ML model does not predict a highly unstable compound to be stable, the ability to correctly predict new stable materials is often more closely related to how accurately a ML model can predict the hull distances of compounds that are stable or nearly stable, i.e. compounds that have hull distances less than ~50 meV/atom. Hull distance predictions of stable/nearly stable compounds are shown in Figure 2 (e) and (f). For this latter set of compounds, iCGCNN hull distance prediction errors are lower by 33% and 28% in terms of MAE and RMSE respectively compared to those of CGCNN. For CGCNN and iCGCNN respectively, MAE measured on the nearly stable compounds are 28% and 16% higher compared to when they are measured on the entire test dataset. The worse predictive accuracy for the stable/nearly stable compounds is likely because there are far fewer stable compounds than there are unstable compounds in the training data (e.g., out of the training set of ~200,000 compounds only 5.1% are stable), making it more difficult to learn the formation energies/hull distances of the stable/nearly stable compounds.

CGCNN and iCGCNN models using Strategy 2 were trained directly on the hull distances of the training entries as queried from the OQMD, and then used to predict the hull distances of the testing entries without having to construct the convex hulls. Model performances for Strategy 2 are summarized in Figure 3. The overall MAE's for the original and improved models are 48.9 and 38.7 meV/atom respectively (Figure 3 (a) (b)). We find that the improved model outperforms the original model by close to 20%. For nearly stable compounds, MAE's for the original and improved models are 61.6 and 50.1 meV/atom respectively (Figure 3 (c) (d)). The difference between the MAE values for stable/nearly stable compounds is close to 20%, consistent with the gap between the overall MAE values. Again, as in Strategy 1, we observe that MAE measured on the stable/nearly stable compounds are higher compared to when they are measured on the entire test dataset by 26% and 29% for CGCNN and iCGCNN respectively.

Comparing Strategies 1 and 2, we find that hull distance prediction errors in general are higher for Strategy 2. For stable/nearly stable compounds, MAE resulting from Strategy 1 was almost 40% and 30% lower for the original and improved CGCNN models respectively. Furthermore, for Strategy 2, the DFT-calculated and ML-predicted hull distances have a negative coefficient of determination ($R^2$) as shown in Figure 3 (c), (d), implying the lack of a linear correlation between the calculated and predicted stabilities. We speculate that there are two causes for this result: First, the hull distance is inherently a more complicated property to learn compared to formation energy. As opposed to formation energy which is calculated with respect to the elemental reference states, hull distance is calculated with respect to the ground state phases. The number of ground states phases that are required to be learned for Strategy 2 is much larger than the number of elemental reference states that needs to be learned for Strategy 1. For example, as opposed to only having to learn the elemental reference states of Li and O to predict the formation energy of $Li_2O$, a ML model must learn about the ground state Li, $Li_2O_2$, $LiO_2$, $LiO_3$, and $O_2$ phases to directly predict the hull distance. Second, convex hulls in the OQMD, or any high-throughput materials database that is currently available, are very much incomplete as there are still potentially thousands of compounds that have not been discovered yet. Thus, for Strategy 2, it is very likely that the training data include incorrect hull distance values derived from incomplete convex hulls. For example, suppose there is a compound with the composition of $Li_4O$ in the binary Li-O space that is stable but is yet to be discovered. If such hypothetical compound does exist, it means that our knowledge of the convex hull in the Li-O space so far has been incomplete, and the convex hull distances that we calculate for compounds with compositions such as $Li_3O$ in the OQMD are incorrect because they are based on a convex hull construction that excludes the stable $Li_4O$ phase. Such incorrect information in the training data may hinder the learning process of the ML models.

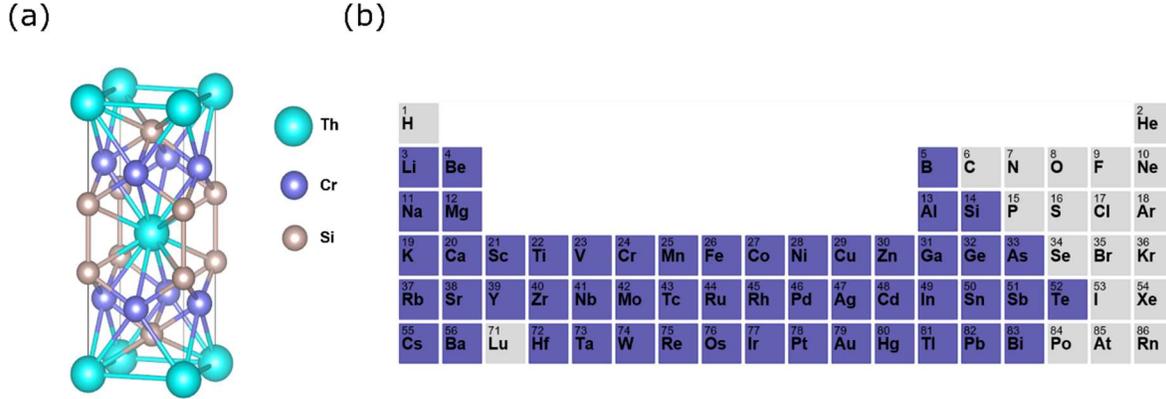

**Figure 4** (a) Structure of ThCr$_2$Si$_2$ (b) Periodic table where colored elements were substituted into ThCr$_2$Si$_2$ structure to generate new compounds.

## Using CGCNN and iCGCNN to accelerate high-throughput DFT searches: Discovering new stable ThCr2Si2-type materials

The ThCr$_2$Si$_2$ structure-type, illustrated in Figure 4 (a), is one of the most commonly observed crystal structures in nature. It is the 5$^{th}$ most common crystal structure among ternary intermetallics [46], accounting for 289 of the 13,026 ternary compounds in the Pearson's Crystal Data [47]. The crystal structure has often been identified with materials that exhibit interesting properties such as superconductivity and valence fluctuation [44]. The number of possible compositions for the ThCr$_2$Si$_2$ structure is about 500,000, and at the time of this study, there were 538 stable ThCr$_2$Si$_2$-type compounds in the OQMD. This implies that if we conducted a undirected high-throughput search for these compounds, we would approximately identify one new stable compound for every 1000 DFT calculations. For a high-throughput DFT search, we could define a success rate as the ratio of number of stable compounds identified to the number of DFT calculations that were performed to identify those stable compounds. The success rate of an undirected high-throughput search for ThCr$_2$Si$_2$-type materials would then be around 0.1%. In this section, we conduct a ML-assisted high-throughput search for materials in the ThCr$_2$Si$_2$ structure-type by using the original and improved CGCNN models in parallel with Strategies 1 and 2 to improve the success rate.

First, new prototype compounds were generated by substituting elements into the original ThCr$_2$Si$_2$ structure. Only metallic elements that are not rare earth, totaling 52 elements, were considered for substitution (Figure 4 (b)), resulting in 132,600 (52x51x50) variations. CGCNN and iCGCNN models that have been trained for the comparative study in the previous section were then used to predict and screen for potentially stable compounds among the newly generated prototype compounds. For the compounds that were predicted to be stable by the ML models, DFT was used to calculate the formation energies and subsequently the hull distances to validate their thermodynamic stability. All DFT calculations were performed within the OQMD framework. Finally, we evaluate the performance of each model based on the number of newly discovered compounds and success rate of their respective high-throughput search.

**Table 1** Performance breakdown of CGCNN and iCGCNN using Strategies 1 and 2 in predicting stable compounds out of the 132,600 new prototype ThCr$_2$Si$_2$-type materials.

|  |  | # of compounds predicted to be stable by ML | # of compounds validated to be stable by DFT* | Success rate % | Avg. hull distance of DFT unstable compounds (meV/atom) |
|---|---|---|---|---|---|
| CGCNN | Strategy 1 | 139 | 35 (33) | 25 | 98.4 |
|  | Strategy 2 | 489 | 62 (80) | 13 | 169.8 |
| iCGCNN | Strategy 1 | 202 | 69 (60) | 34 | 94.6 |
|  | Strategy 2 | 315 | 112 (113) | 35 | 71.8 |

(* numbers in parenthesis indicate the number of compounds that are not stable but are nearly stable as calculated by DFT)

The results of the search are summarized in Table 1. Combining the compounds predicted to be stable from both Strategies 1 and 2 without overlap, the original CGCNN predicted a total of 556 unique compounds to be stable, out of which 72 were confirmed to stable through DFT. Out of the 423 unique compounds predicted to be stable by iCGCNN, 133 were confirmed to be stable through DFT, nearly twice the number of stable compounds identified by CGCNN. The 31% success rate of iCGCNN is a factor of 2.4 greater than the 13% success rate of the original CGCNN and a factor of 310 greater than the 0.1% success rate of an undirected high-throughput search. Further investigation of the compounds that were not stable as calculated by DFT revealed that the number of nearly stable compounds found by iCGCNN was 82% and 41% higher than that of CGCNN for Strategies 1 and 2 respectively. The average hull distances of the compounds predicted by iCGCNN that were DFT unstable were also lower by 3.9% and 57.7% for Strategies 1 and 2 respectively. Both results show that iCGCNN predicted compounds are consistently closer to the convex hull than the compounds predicted by CGCNN. Overall, iCGCNN is far more accurate and efficient in discovering new stable compounds than CGCNN.

Through this survey, we performed a total of 757 DFT calculations and identified 143 stable unique compounds, of which 97 compounds have not yet been reported in literature to the best of our knowledge. Among the 97 compounds, 42 are significantly stable with hull distances less than -50 meV/atom. While all 97 compounds identified in this survey are computational predictions that await experimental validation, compounds with the significantly negative convex hull distances are most likely to be synthesizable, and hence should be prioritized in any experimental synthesis effort. All 97 compounds are listed by the convex hull distance in Supplementary Table S1. The number of stable occurrences for each element on the Th-, Cr-, and Si- site are shown in Figure 5. The Th- site, which has the highest coordination among the 3 sites, is mostly occupied by alkaline elements that generally have large radii. The Cr- site and Si- site are mostly occupied by transition metals and metalloids respectively. Finally, we emphasize that we have discovered 97 potentially new stable compounds by only performing 757 DFT calculations, a success rate of 13% that implies that we have accelerated the high-throughput search for $ThCr_2Si_2$-type materials by a factor 130 using both CGCNN and iCGCNN.

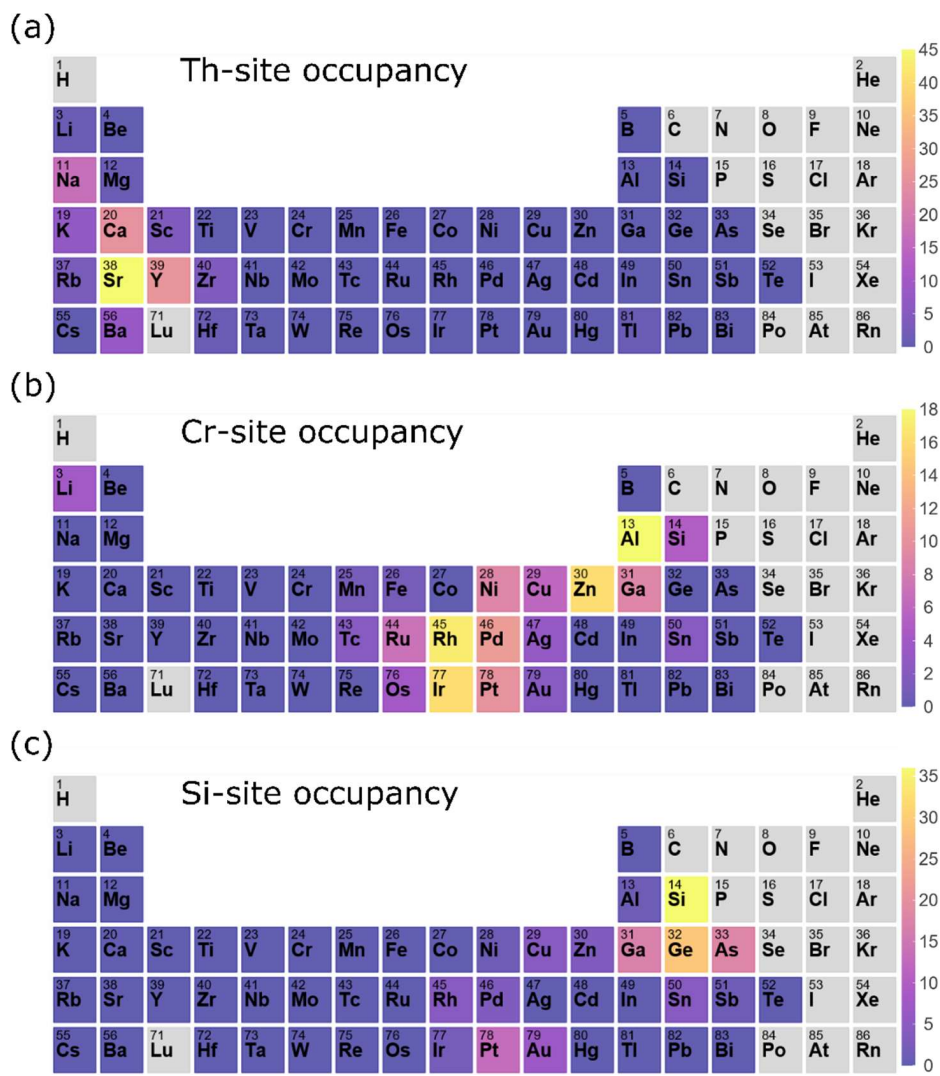

**Figure 5** Periodic tables where elements are color-coded based on the number of stable occurrences on the (a) Th- site, (b) Cr- site, and (c) Si- site.

## Conclusions

The CGCNN model provides a highly accurate and flexible ML framework in which material descriptors are adaptively extracted according to the task at hand and thus, allows us to bypass the painstaking process of handcrafting the material descriptors ourselves. We have presented an improved CGCNN model (iCGCNN) to demonstrate that this framework can be further improved. This was done by integrating the Voronoi tessellation information of the crystal structure, explicitly encoding the 3-body correlations of neighboring constituent atoms, and optimizing the chemical representation of interatomic bonds in the crystal graphs. We trained and tested the original and improved CGCNN models on OQMD data to compare their hull distance predictive accuracy using two approaches: 1) predicting the formation energy and subsequently calculating the hull distance relative to OQMD constructed convex hull and 2) bypass the need to construct the convex hull by directly predicting the hull distance. In both approaches, there were significant gaps between the predictive accuracies, where the iCGCNN performed 25% and 20% better than CGCNN for the former and latter approach respectively in terms of the MAE measured on the entire testing data. Finally, when used to predict new stable compounds with $ThCr_2Si_2$-structure, iCGCNN not only identified twice as many more stable compounds than the original CGCNN, it exhibited a success rate that was greater by a factor 2.4. Beyond comparing the two ML models, we discovered 97 new stable compounds in a high-throughput search that was accelerated by a factor of 130 using both CGCNN and iCGCNN. Its excellent performances in screening stable compounds suggests iCGCNN can be used to greatly accelerate materials discovery.

## Methods

### Data

We use DFT-calculated thermodynamic data from the OQMD for training, validating, and testing ML models throughout this work. OQMD v1.1 contains about ~450,000 DFT-calculations of unique ordered inorganic compounds, including ~40,000 experimentally known ones from the Inorganic Crystal Structure Database (ICSD) [39, 40], and the rest hypothetical ones generated from commonly occurring structural prototypes. All DFT calculations in the OQMD are performed with the Vienna *Ab Initio* Simulation Package (VASP) [41, 42]. The details of the methodology and settings used for the high-throughput calculations are explained in Ref. [3]. All ML models in this work are trained on a set of ~180,000 compounds and validated on another ~20,000 compounds, all randomly selected from the OQMD with no overlap. Models are tested on a separate set of ~230,000 compounds that are not included in the training or validation data [2, 3].

## Acknowledgements

This work was performed under the following financial assistance Award 70NANB14H012 from U.S. Department of Commerce, National Institute of Standards and Technology as part of the Center for Hierarchical Materials Design (CHiMaD). The authors also acknowledge the financial support of Toyota Research Institute (TRI). This work used the Extreme Science and Engineering Discovery Environment (XSEDE), which is supported by National Science Foundation grant number ACI-1548562. This work was also supported in part through the computational resources and staff contributions provided for the Quest high performance computing facility at Northwestern University which is jointly supported by the Office of the Provost, the Office for Research, and Northwestern University Information

Technology. The authors thank Vinay Hegde, Eric Isaacs, Mohan Liu, Sean Griesemer, Abhijith Gopakumar, and Koushik Pal for helpful discussion.

## Author Contributions

CW conceived the project, and jointly developed the method with CWP. CWP wrote all software and performed the necessary calculations with help and guidance from CW. CWP led the manuscript writing, with contributions from all other authors.

## Competing interests

The authors declare no conflict of interest